\documentstyle[preprint,revtex,eqsecnum]{aps}
\begin{document}
\draft
\begin{title}
Phase diagram of the $S=\frac{1}{2}$ quantum spin chain \\
with bond alternation
\end{title}
\author{M. Yamanaka, Y. Hatsugai\cite{address} and M. Kohmoto}
\begin{instit}
Institute for Solid State Physics, University of Tokyo \\
7-22-1, Roppongi, Minato-ku, Tokyo 106 Japan
\end{instit}
\begin{abstract}

We study the ground state properties
of the bond alternating $S=1/2$ quantum spin chain
 whose Hamiltonian is

\begin{equation}
H=\sum_j (S_{2j}^x S_{2j+1}^x +S_{2j}^y S_{2j+1}^y
+\lambda S_{2j}^z S_{2j+1}^z )
+\beta \sum_j {\bf S}_{2j-1} \cdot {\bf S}_{2j}
\ .
\end{equation}
When $\beta=0$, the ground state is a collection of local singlets
with a finite excitation gap.
In the limit of strong ferromagnetic coupling $\beta \to - \infty$,
this is equivalent to the $S=1 \ XXZ$ Hamiltonian.
It has several ground state phases in the $\lambda$-$\beta$ plane
including the gapful Haldane phase.
They are characterized by a full breakdown,
partial breakdowns and a non-breakdown of the hidden discrete
$Z_2 \times Z_2$ symmetry.
The ground state phase diagram is obtained by series expansions.

\end{abstract}
\narrowtext

\section{Introduction
}
\label{sec:intro}

We study the ground state properties
of an $S=1/2$ alternating spin chain
with the Hamiltonian

\begin{equation}
H=\sum_j (S_{2j}^x S_{2j+1}^x +S_{2j}^y S_{2j+1}^y
+\lambda S_{2j}^z S_{2j+1}^z )
+\beta \sum_j {\bf S}_{2j-1} \cdot {\bf S}_{2j}
,
\label{s2}
\end{equation}
where ${\bf S}_j=(S^x_j,S^y_j,S^z_j)$ are $S=1/2$ spin operators.
This model has two exchange couplings $1$ and $\beta$ alternately.
When $\beta=1$ and $\lambda=1$,
it is the spherically symmetric $S=1/2$ Heisenberg model
which is solvable by the Bethe ansatz \cite{Bethe}.
When $\beta=0$, the ground state is composed of local singlets
that are formed by spins on sites $2j$ and $2j+1$.
In the limit of strong ferromagnetic coupling $\beta \to - \infty$,
spins on sites $2j-1$ and $2j$ form a local triplet.
The first order degenerate perturbation in $1/\beta$ from this limit
gives the $S=1$ spin chain with the Hamiltonian

\begin{equation}
H^{S=1}=\sum_i ( \ S_i^x S_{i+1}^x +S_i^y S_{i+1}^y
        +\lambda S_i^z S_{i+1}^z ),
\label{s1}
\end{equation}
where $S^x_i,S^y_i$ and $S^z_i$ are $S=1$ spin operators at site $i$.
When $\lambda=1$, the Hamiltonian (\ref{s1})
is the Heisenberg antiferromagnetic chain with $S=1$.

In $1983$ Haldane conjectured
that there are qualitative differences between integer
and half-integer spin quantum antiferromagnetic chains
based on the large-$S$ expansion \cite{Hald}.
While there is no excitation gap when $S$ is a half-integer
\cite{gapless},
he argued that the Heisenberg Hamiltonian
has a unique disordered ground state with a finite excitation gap
when the spin $S$ is an integer.
These claims have been confirmed
by experimental \cite{exp},
numerical \cite{num0,num,boundaryNH,num1}
and analytical studies \cite{theo} at least for the $S=1$ case.
The elementary excitations and the temperature dependence
of the Haldane phase
are investigated using the defect approach \cite{defect}.
Although the ground state is disordered in the sense
that the usual spin-spin correlation function decays exponentially,
it has turned out that the ground state has
a nontrivial hidden order.
Den Nijs and Rommelse \cite{DN1} argued
that the Haldane phase is characterized
by the hidden antiferromagnetic order
based on the analogy of the preroughning transition.
It is measured by the string order parameter.
Kennedy and Tasaki \cite{KeT} introduced
a nonlocal unitary transformation
which reveals the relation between the hidden antiferromagnetic order
and the hidden discrete $Z_2\times Z_2$ symmetry breaking.
The hidden $Z_2 \times Z_2$ symmetry is completely broken
in the Haldane phase.
This is confirmed numerically \cite{GA}.
Extensions to higher integer $S$ cases
were discussed by several authors \cite{T}.

The $S=1/2$ spin chains with bond alternation
have been investigated by several groups \cite{alt}.
Recently Hida studied an $S=1/2$ Heisenberg chain
with alternating ferromagnetic and antiferromagnetic couplings.
Its Hamiltonian is (\ref{s2}) with $\lambda=1$.
Hida considered a string order parameter
for $S=1/2$ which leads to
the den Nijs-Rommelse string order parameter
in the $\beta \to -\infty$ limit.
He concluded numerically
that the gap and the string order parameter for $\beta=0$
remain non-vanishing in the $\beta \to -\infty$ limit.
It suggests that the $S=1/2$ spin chains with disordered ground state
are characterized by the string order parameters,
and that for $\lambda =1$ there is no phase transition
between $\beta =0$ ($S=1/2$ local singlet-triplet gap)
and $\beta =-\infty$ ($S=1$ Haldane gap),
thus they belongs to the same phase \cite{Hi1,Hi2,Hi3}.

There are studies on the $S=1/2$ spin chains with bond alternation
using other types of nonlocal unitary transformation
\cite{KoT,s2nut}.
In particular, Kohmoto and Tasaki \cite{KoT} studied
the Hamiltonian (\ref{s2}) by the nonlocal unitary transformation
used in the work of the Ashkin-Teller model
by Kohmoto, den Nijs and Kadanoff \cite{KdK}.
They found that this transformation plays a role
of the Kennedy-Tasaki unitary transformation for the $S=1$ chain,
although they are distinct.
The nonlocal unitary transformation maps
the $S=1/2$ alternating chain
to an $S=1/2$ quantum spin system which is similar
to the highly anisotropic version (one-dimensional quantum system)
of the two-dimensional Ashkin-Teller model.
It also maps the string order parameters of the original chain
to the ferromagnetic ``local'' order parameters
in the transformed system.
They argued the correspondence between the breaking
of the $Z_2\times Z_2$ symmetry in the transformed system
and the hidden antiferromagnetic order in the original system.
In the transformed system,
they showed that the region including the decoupled model ($\beta=0$)
and the $S=1$ chain are characterized
by a full breakdown of the $Z_2 \times Z_2$ symmetry.
The N\'eel ordered phase (in terms of $S=1$)
is characterized by a partial breakdown.
They also proposed a ground state phase diagram.

The unitary transformation maps the quantum spin chain
with bond alternation into the Ashkin-Teller type quantum spin system
on a pair of chains.
The spin chains are similar to the Ashkin-Teller quantum chain.
The latter model is obtained by
a highly anisotropic limit of the two-dimensional Ashkin-Teller model
\cite{KdK,KaK}.
The two-dimensional Ashkin-Teller model \cite{AT}
consists of two Ising models on a square lattice
coupled by a four-spin interaction.
In the time-continuum Hamiltonian formalism,
a two-dimensional classical system is reduced
to a one-dimensional quantum system
by taking the extreme lattice anisotropic limit.
The transfer-matrix method is used
to convert a statistical mechanics problem at a finite temperature
in two dimensions into a ground state problem
for one-dimensional quantum Hamiltonian \cite{Kog2}.
In this formalism, the ground state energy of a quantum system
is the free energy of the corresponding two-dimensional
classical system.
The excitation gap in the quantum system corresponds
to the inverse of the correlation length of the classical system.
If there is no excitation gap in the quantum system,
the classical system is critical.
Namely phase boundaries of the ground states phase diagram
of a one-dimensional quantum system correspond to critical points
of the two-dimensional classical spin system.
Thus we shall use the concepts of the critical phenomena
(universality, critical indices etc.)
in order to discuss the ground state phase diagram
of the one-dimensional quantum system even in the parameter region
where a corresponding two-dimensional classical system
does not exist.

The purpose of the present paper is to determine
the phase diagram of the Hamiltonian (\ref{s2}) quantitatively
by a series expansion technique.
Some limiting cases are considered analytically.
The critical lines and critical indices are evaluated
from the Pad\'e method \cite{Pade}.
The obtained phase diagram shows a rich structure.
It contains the gapful Haldane phase, the N\'eel phases,
the $XY$-like gapless phase.
It also has a critical line
with continuously varying critical indices.

\section{the disorder operators and the $Z_2 \times Z_2$ symmetry
breakings}
\label{sec:disorder}

We shall perform series expansions in terms of $\beta$.  The
unperturbed system ((\ref{s2}) with $\beta =0$) has the ground state
which is a collection of uncorrelated local singlets.  It is
disordered in the sense that the expectation values of local order
parameters vanish and series expansions cannot be applied to these
quantities. Thus we consider the disorder operators which are nonzero
in disordered phases and zero in an ordered phase \cite{KW}.  We
choose the following disorder operators

\begin{equation}
D^{xy} (j)
= \prod_{k>j}
\{ \ -2( S_{2k}^x  S_{2k+1}^x +S_{2k}^y  S_{2k+1}^y )\ \}
= \prod_{k>j} \{ \ -( S_{2k}^+  S_{2k+1}^-
+S_{2k}^-  S_{2k+1}^+ )\ \},
\label{diso1}
\end{equation}
and

\begin{equation}
D^{z} (j) = \prod_{k>j} (\ -4S_{2k}^z  S_{2k+1}^z \ ),
\label{diso2}
\end{equation}
to characterize the ground states of the Hamiltonian (\ref{s2}),
where $S^{\alpha}$ are the $S=1/2$ spin operators.

We describe the derivation of above disorder operators
and their relations to the hidden $Z_2 \times Z_2$ symmetry.
The nonlocal unitary transformation reveals
the $Z_2 \times Z_2$ symmetry of the Hamiltonian (\ref{s2}).
This transformation is exactly the same as that used
by Kohmoto, den Nijs and Kadanoff \cite{KdK},
which maps the staggered $XXZ$ model
into the Ashkin-Teller quantum chain.
Applying this transformation to the Hamiltonian (\ref{s2}),
we get

\begin{equation}
H=-\sum_i (\sigma_i^x + \tau_i^x + \lambda \sigma_i^x \tau_i^x)
-\beta \sum_i (\sigma_i^z \sigma_{i+1}^z + \tau_i^z \tau_{i+1}^z
+ \sigma_i^z \sigma_{i+1}^z \tau_i^z \tau_{i+1}^z),
\label{hamat1}
\end{equation}
where $\sigma_i^{\alpha}$ and $\tau_i^{\alpha}$
($\alpha$ = $x$, $y$ and $z$) are different kinds of Pauli matrices
at site $i$.
For details, see Appendix B of Ref. 21.

This Hamiltonian has symmetries of the dihedral group of order $4$
which includes several $Z_2$ symmetries.
There are many varieties of order parameters
to measure the spontaneous symmetry breaking of these symmetries.
We choose the following order operators:

\begin{equation}
O_{\pm}=\frac{\sigma_i^z + \tau_i^z}{2},
\label{atorder1}
\end{equation}
\begin{equation}
O_2=\sigma_i^z \tau_i^z.
\label{atorder2}
\end{equation}

The disorder operators which are dual to these order operators are

\begin{equation}
D_{\pm}(i)=\prod_{k>i} \frac{\sigma_k^x + \tau_k^x}{2},
\label{diso5}
\end{equation}
\begin{equation}
D_{2}(i)=\prod_{k>i} \sigma_k^x \tau_k^x,
\label{diso6}
\end{equation}
respectively.
These disorder operators have nonzero expectation values
in disordered phases.
Applying the inverse of the above transformation
to the disorder operators
(\ref{diso5}) and (\ref{diso6}),
we get the disorder operators (\ref{diso1}) and (\ref{diso2})
of the Hamiltonian (\ref{s2}).

A nonzero expectation value of one of the disorder operators
corresponds to a spontaneous breaking of a $Z_2$ symmetry.
The ground state phases are characterized by a full breaking,
partial breakings and nonbreakings of the $Z_2 \times Z_2$ symmetry.
Thus the ground state phase diagram can be obtained
from analysis of the disorder operators (\ref{diso1})
and (\ref{diso2}).

The disorder operators (\ref{diso1}) and (\ref{diso2})
are related to the string order parameters
of den Nijs and Rommelse, and Hida.
It is expected that they have the same critical properties.

\section{Phase diagram}
\label{sec:phasedia}

Before going to the detailed analysis, we summarize our results
by showing the obtained phase diagram in Fig.\ \ref{phase}.
This phase diagram have seven phases and eight critical lines.\\
(1) The gapful Haldane phase (A).
It is characterized by
$\langle \ D^{\alpha} \ \rangle \ne 0$,
($\alpha = xy$ or $z$).
\\
(2) The $S=1$ N\'eel phase (B) which is characterized by
$\langle \ D^{z} \ \rangle \ne 0$
and $\langle \ D^{xy} \ \rangle =0$.
This region continuously connects to the N\'eel phase of the $S=1$
Hamiltonian (\ref{s1}).\\
(3) The $XY$-like gapless phase (C) where
$\langle \ D^{\alpha} \ \rangle$ ($\alpha=xy$ and $z$) vanishes.
The correlation function decays algebraically in the entire region
and the critical indices vary continuously.\\
(4) The ferromagnetic phase (D).\\
(5) The disordered phase (E) in which
the ground state is a disordered dimer state
with $\langle \ D^{\alpha} \ \rangle$ =$0$ ($\alpha=xy$ and $z$).\\
(6) The $S=1/2$ N\'eel ordered phase (F).
This phase is characterized by
$\langle \ D^{z} \ \rangle \ne 0$
and $\langle \ D^{xy} \ \rangle =0$. \\
(7) The N\'eel ordered phase (G)
where spins on sites $2j$ and $2j+1$ prefer to align parallel
and those on sites $2j-1$ and $2j$ prefer to align antiparallel.

These phases are separated by the following critical lines. \\
(1) Line 1 is expected to be in the Gaussian model universality class
\cite{KB}.
Critical indices vary continuously on this line
for $-1 < \lambda <1$.
This critical line is expected to bifurcate
at $\lambda =1$ \cite{Kada}.
\\
(2) Line 2 and Line 3 are expected to be
in the Ising model universality class.
Line 2 approaches the line $\beta =2$ for large $\lambda$.
Line 3 approaches the line $\beta = \frac{1}{2} \lambda$
for large $\lambda$.
\\
(3) Line 4 is the phase boundary
between the Haldane phase and the N\'eel phase
(in terms of the $S=1$ model)
and belongs to the Ising model universality class.\\
(4) Line 5 is the phase boundary between the $XY$-like gapless phase
and the Haldane phase, and is expected to be in the universality
class of the Kosterlitz-Thouless transition \cite{KT}.\\
(5) Line 6 ($\lambda=-1$, $\beta<0$) is the phase boundary
between the $XY$-like phase and the ferromagnetic phase,
and is expected to be in the universality class
of the KDP transition. \cite{KDP}\\
(6) The Hamiltonian (\ref{s2}) is solvable on line $\beta =0$.
The ground state is degenerate on Line 7.
Thus, Line 7 is a critical line \cite{KdK}.
\\
(7) Line 8 belongs to the Ising model universality class
and approaches the line $\beta =-\frac{1}{2} \lambda$,
for large $\vert \lambda \vert$. \\

\subsection{Limiting Cases}

In this subsection we discuss the phase diagram in several limits.
Some of the critical lines are obtained by the duality.

\subsubsection{limit\ $\lambda \gg \vert \beta \vert \sim 1$}

Since the term with $\lambda$ dominates,
we can restrict ourselves to the space spanned by

\begin{equation}
\vert \ \{ \alpha_j \} \ \rangle
= \bigotimes_j \ \vert \ \alpha_j \ \rangle_{2j,2j+1}
\ ,
\label{state0}
\end{equation}
where the state $\vert \alpha_j \rangle_{2j,2j+1}$ is defined by
\begin{equation}
\vert \ \ \alpha_j =\uparrow \ \ \rangle_{2j,2j+1}
=\vert \uparrow \rangle_{2j} \ \vert \downarrow \rangle_{2j+1},
\ \ \ \ \
\vert \ \ \alpha_j =\downarrow \ \ \rangle_{2j,2j+1}
=\vert \downarrow \rangle_{2j} \ \vert \uparrow \rangle_{2j+1}.
\label{state1}
\end{equation}
Here, $\vert \uparrow \rangle_j$ and $\vert \downarrow \rangle_j$
are eigenstates of $S_j^z$ with eigenvalues $+1/2$
and $-1/2$ respectively.
We treat the other parts of the Hamiltonian
by the degenerate perturbation theory
within the space spanned by (\ref{state0}).
The first order perturbation gives the effective Hamiltonian

\begin{equation}
H=\frac{1}{2} \sum_i \sigma_i^x
- \frac{1}{4} \beta \sum_i \sigma_i^z \sigma_{i+1}^z  \label{eff1}
\ ,
\end{equation}
where $\sigma_i^{\alpha}$ are the Pauli matrices which operate
on $\vert \alpha_i \ \rangle_{2i,2i+1}$.
This is the highly anisotropic version
of the two-dimensional Ising model \cite{Kog2}.
This Hamiltonian have the property

\begin{equation}
-\frac{\beta}{2} H \left( \frac{4}{\beta} \ ,
\ \widetilde{\sigma^{\alpha}} \right)
= H (\beta , \sigma^{\alpha} )
\ ,
\label{dualham}
\end{equation}
under the dual transformation

\begin{equation}
\widetilde{\sigma_i^z} = \prod_{k=1}^i \sigma_k^x ,\ \ \ \
\widetilde{\sigma_i^x} = \sigma_i^z \sigma_{i+1}^z \ .
\label{duality2}
\end{equation}
For $\beta <0$, this duality of the Ising model tells us
that $\beta =-2$ is the critical line.
Thus the critical line between the Haldane phase
and N\'eel phase approaches
$\beta =-2$ as $\lambda$ becomes large.
Above this line (phase A in Fig.\ \ref{phase}),
the $Z_2 \times Z_2$ symmetry is fully broken.
Below this line (phase B in Fig.\ \ref{phase}),
the ground state breaks half of the $Z_2 \times Z_2$ symmetry.
Critical properties corresponding to a partial breakdown
of $Z_2 \times Z_2$ symmetry is considered
as the Ising model universality class.
Thus Line 4 belongs to the Ising model universality class.

For $\beta >0$, we apply a transformation
$S_{2j}^x \ \to \ -S_{2j}^x$,
$S_{2j}^y \ \to \ -S_{2j}^y$
and
$S_{2j}^z \ \to \ S_{2j}^z$
only for spins on the $2j$ sites.
Then we get the same duality relation as (\ref{dualham})
except for the sign of $\beta$.
It implies that $\beta=2$ is a self-dual line
for the effective Hamiltonian.
Thus Line 2 which approaches $\beta=2$ as $\lambda$ becomes large
belongs to the Ising model universality class.

\subsubsection{limit \ $\beta,\ \lambda \gg 1$}

The Hamiltonian is approximated by

\begin{equation}
H =\lambda \sum_j S_{2j}^z S_{2j+1}^z
+ \beta \sum_j {\bf S}_{2j-1} \cdot {\bf S}_{2j}
\ .
\label{eff2}
\end{equation}
When $\lambda$ is sufficiently large,
the $\lambda$-coupling dominates in the Hamiltonian (\ref{eff2})
and spins on sites $2j$ and $2j+1$ must form a state
$\vert \uparrow \rangle_{2j} \vert \downarrow \rangle_{2j+1}$
or $\vert \downarrow \rangle_{2j} \vert \uparrow \rangle_{2j+1}$.
In this bases, spins on sites $2j-1$ and $2j$ favors the states,
$\vert \uparrow \rangle_{2j-1} \vert \downarrow \rangle_{2j}$
or $\vert \downarrow \rangle_{2j-1} \vert \uparrow \rangle_{2j}$,
due to the antiferromagnetic coupling $\beta$.
Therefore, the ground state is given either by
$\bigotimes_j (\vert \uparrow \rangle_{2j} \
\vert \downarrow \rangle_{2j+1} )$
or
$\bigotimes_j (\vert \downarrow \rangle_{2j} \
\vert \uparrow \rangle_{2j+1} )$.
It is a doubly degenerate N\'eel ordered state.
The energy per site is $\frac{1}{8} \lambda -\frac{1}{8} \beta $.
On the other hand,
spins on sites $2j-1$ and $2j$ favor to form singlet
for $\beta \gg \lambda$.
The ground state is given by an array of them.
It is disordered and the energy is $-\frac{3}{8} \beta$.

To investigate the phase boundary between these phases,
we map the Hamiltonian (\ref{eff2}) into the model
without bond alternation.
Applying the nonlocal unitary transformation
(see Sec.\ \ref{sec:disorder}) to the Hamiltonian (\ref{eff2}),
we get (see (\ref{hamat1}))

\begin{equation}
H=- \lambda \sum_i \sigma_i^x \tau_i^x
 -\beta \sum_i (\sigma_i^z \sigma_{i+1}^z + \tau_i^z \tau_{i+1}^z
 + \sigma_i^z \sigma_{i+1}^z \tau_i^z \tau_{i+1}^z).
\label{hamat2}
\end{equation}
We map the space spanned by the eigenstates of $\sigma_i^z$
and $\tau_i^z$
to that spanned by the eigenstates of $\sigma_i^z \tau_i^z$
and $\tau_i^z$.
We define the operator $\widehat{\tau_i^{z}}$
by $\widehat{\tau_i^{z}} =\sigma_i^z \tau_i^z$.
Using this notation, the Hamiltonian (\ref{hamat2}) becomes

\begin{eqnarray}
H & = & -\lambda \sum_i \tau_i^x
 -\beta \sum_i (\widehat{\tau_i^z} \widehat{\tau_{i+1}^z}
               +\tau_i^z \tau_{i+1}^z
 +\widehat{\tau_i^z} \tau_i^z \widehat{\tau_{i+1}^z} \tau_{i+1}^z) \\
  & = & -\lambda \sum_i \tau_i^x
 -\beta \sum_i \{ (1 +\tau_i^z \tau_{i+1}^z)
              \widehat{\tau_i^z} \widehat{\tau_{i+1}^z}
               +\tau_i^z \tau_{i+1}^z \}.
\label{hamat3}
\end{eqnarray}
In this Hamiltonian,
the coupling of $\widehat{\tau_i^z}$ is negative,
since $1 +\tau_i^z \tau_{i+1}^z$ is always positive.
Thus $\widehat{\tau}$ spins appear as the one-dimensional Ising model
with the ferromagnetic coupling.
In the ground state,
$\widehat{\tau}$ spins are completely ordered ferromagnetically.
We can replace the operator
$\widehat{\tau_i^z} \widehat{\tau_{i+1}^z}$
by its expectation value
$\langle \widehat{\tau_i^z} \widehat{\tau_{i+1}^z} \rangle =1$.
Thus we have

\begin{equation}
H=-\lambda  \sum_i \tau_i^x
 -2\beta \sum_i \tau_i^z \tau_{i+1}^z,
\label{hamat4}
\end{equation}
apart from a trivial additive constant.
The effective Hamiltonian (\ref{hamat4}) is the Ising model
in a transverse magnetic field.
Therefore, the boundary of the phases is $\lambda
= \frac{1}{2} \beta$
in this limit (Line 3 in Fig\ \ref{phase}).
Note that the phase transition for $\tau$ spins
across critical line 3.
Since the Hamiltonian (\ref{hamat2}) is symmetric with respect to
$\tau$ spins and $\sigma$ spins,
$\sigma$ spins have the same behavior as $\tau$ spins.
Therefore, the order parameter (\ref{atorder1})
has nonzero expectation value in Phase E.
The disorder operator whose series has singularity on Line 3
is $D^z$.
Below this line (phase F in Fig\ \ref{phase}), the system is ordered
in terms of $S=1/2$ N\'eel order and is
characterized by $\langle \ D^{z} \ \rangle \ne 0$
and $\langle \ D^{xy} \ \rangle =0$.
Thus the ground state breaks half of the $Z_2 \times Z_2$ symmetry.
Above this line (phase E in Fig\ \ref{phase}) ,
the system is disordered
and is characterized by $\langle \ D^{\alpha} \ \rangle =0$
($\alpha=xy$ and $z$).
The $Z_2 \times Z_2$ symmetry is fully restored.
Line 3 belongs to the Ising model universality class,
since the $Z_2$ symmetry is broken across this line.

\subsubsection{
limit \ $\beta ,\ \vert \lambda \vert \ (\lambda<0) \gg 1$}

Similar to the previous subsection,
the Hamiltonian of this regime is

\begin{equation}
H=-\vert \lambda \vert \sum_j S_{2j}^z S_{2j+1}^z
+ \beta \sum_j {\bf S}_{2j-1} \cdot {\bf S}_{2j} \ .
\end{equation}
In the case of large $\vert \lambda \vert$,
spins on sites $2j$ and $2j+1$
prefer to form a state
$\vert \uparrow \rangle_{2j} \vert \uparrow \rangle_{2j+1}$
or $\vert \downarrow \rangle_{2j} \vert \downarrow \rangle_{2j+1}$.
Spins on sites $2j-1$ and $2j$ favor the state
$\vert \uparrow \rangle_{2j} \vert \downarrow \rangle_{2j+1}$
or $\vert \downarrow \rangle_{2j} \vert \uparrow \rangle_{2j+1}$,
since the coupling $\beta$ is antiferromagnetic.
Therefore, the ground state is given by
$\bigotimes_j \ \left( \ \vert \uparrow \rangle_{4j} \
\vert \uparrow \rangle_{4j+1} \ \vert \downarrow \rangle_{4j+2} \
\vert \downarrow \rangle_{4j+3} \ \right) $
and
$\bigotimes_j \ \left( \ \vert \downarrow \rangle_{4j} \
\vert \downarrow \rangle_{4j+1} \ \vert \uparrow \rangle_{4j+2} \
\vert \uparrow \rangle_{4j+3} \ \right) $.
The energy of this state is
$-\frac{1}{8} \vert \lambda \vert -\frac{1}{8} \beta$ per site.
For $\beta \gg \vert \lambda \vert$,
spins on sites $2j-1$ and $2j$ favor a singlet state.
The ground state is given by a collection of local singlet
with the energy
$-\frac{3}{8} \beta$.
In this case, the phase boundary is the line
$\lambda =-\frac{1}{2} \beta$ (Line 8 in Fig.\ \ref{phase}).
Below this line (Phase G in Fig.\ \ref{phase}),
the system breaks half of the $Z_2 \times Z_2$ symmetry.
Thus the transition across this line belongs
to the Ising model universality class.

\subsubsection{line \ $\beta=0$}

We have the simplest situation.
The Hamiltonian is written

\begin{equation}
H=\sum_j ( S_{2j}^x S_{2j+1}^x + S_{2j}^y S_{2j+1}^y
+ \lambda S_{2j}^z S_{2j+1}^z ) \ .
\end{equation}
It is a collection of independent two-spin systems.
For $\lambda >-1$, the ground state is given by
$\bigotimes_j \{ \frac{1}{\sqrt{2}}
\left( \vert \uparrow \rangle_{2j} \vert \downarrow \rangle_{2j+1}
\ - \ \vert \downarrow \rangle_{2j} \vert \uparrow \rangle_{2j+1}
\right) \} $
with an energy $- \frac{2+\lambda}{8}$ per site.
For $\lambda <-1$, the ground state of an independent two-spin system
is given by
$\vert \uparrow \rangle_{2j} \vert \uparrow \rangle_{2j+1}$
or
$\vert \downarrow \rangle_{2j} \vert \downarrow \rangle_{2j+1}$.
Therefore, the ground state of the total system is
$2^{M/2}$ fold degenerate,
where $M$ is the number of sites.
Therefore, this is a critical line \cite{KdK}.
The energy is $\frac{\lambda}{8}$ per site.

\subsubsection{
the ferromagnetic phase \ $\beta < 0$,\ $\lambda < -1$
}

In this region,
the Hamiltonian is written

\begin{equation}
H =  \sum_j (S_{2j}^x S_{2j+1}^x + S_{2j}^y S_{2j+1}^y
              - \vert \lambda \vert S_{2j}^z S_{2j+1}^z )
    - \vert \beta \vert \sum_j {\bf S}_{2j-1} \cdot {\bf S}_{2j} \ .
\end{equation}
The ground state is given by
$\bigotimes_j \vert \uparrow \rangle_{2j}
\vert \uparrow \rangle_{2j+1}$
or
$\bigotimes_j \vert \downarrow \rangle_{2j}
\vert \downarrow \rangle_{2j+1}$
with the energy $-\frac{\vert \lambda \vert + \vert \beta \vert}{8}$.

Line 6 is expected to be a critical line in the universality class
of the KDP transition \cite{KDP},
since the system is in the perfectly ferromagnetic ordered state
in the $\lambda < -1$ side
and is in the $XY$-like gapless phase
in the $\lambda > -1$ side.

\subsubsection{limit \ $\beta \to -\infty$}

In the $\beta \to -\infty$ limit, the $S=1/2$ Hamiltonian (\ref{s2})
is equivalent to the $S=1$ $XXZ$ Hamiltonian (\ref{s1}).
It has the N\'eel, the Haldane, the $XY$-like
and the ferromagnetic phases
depending on the parameter $\lambda$.
We denote $\lambda_1$ for the boundary
between the N\'eel and the Haldane phases,
$\lambda_2$ for that between the Haldane and the $XY$-like phases,
and $\lambda_3$ for the $XY$-like and the ferromagnetic phases.

The phase boundary between the N\'eel and the Haldane phases
is estimated to be
$\lambda_1 \sim 1.2$ numerically \cite{num0,boundaryNH}.
The critical line line 4 is expected
to approach $\lambda =\lambda_1$ \cite{KoT}.
The boundary between the Haldane and the $XY$-like phase is estimated
$\lambda_2 \sim 0$ numerically \cite{num0,num1}.
The $XY$-like phase boundary
which starts from the point $(\lambda, \beta)
=(-1,0)$ approaches $\lambda = \lambda_2$.

For $\lambda=-1$, the $S=1$ $XXZ$ Hamiltonian (\ref{s1})
can be mapped
to the ferromagnetic Heisenberg model by the rotation
in the spin space for the $2j$ sites,
$S_{2j}^x \to -S_{2j}^x$, $S_{2j}^y \to -S_{2j}^y$
and $S_{2j}^z \to S_{2j}^z$.
The ground state have degeneracy due to
the rotational symmetry at $\lambda =-1$.
On the other hand, for $\lambda <-1$,
the ground state is $2$ fold degenerate.
Therefore, the phase boundary between the $XY$-like
and the ferromagnetic phases is $\lambda =-1$ in this limit.

\section{Series expansion}
\label{sec:levela}

We will make a series expansion in terms of $\beta$
and obtain series
of the ``specific heat'', the ``magnetization''
and the ``susceptibility''
to estimate critical points and critical indices.

\subsection{Outline of the Method}
\label{method1}

It is necessary to have long series
to extract reliable estimates of the critical points
and critical indices.
We use the linked cluster expansion method proposed
by Kadanoff and Kohmoto \cite{KK}.
It is suited to calculate high order terms effectively.

Hamiltonian (\ref{s2}) is written

\begin{equation}
H=T+\beta V \ ,
\label{ham}
\end{equation}
where

\begin{equation}
T=\sum_j ( S_{2j}^x S_{2j+1}^x + S_{2j}^y S_{2j+1}^y
+ \lambda S_{2j}^z S_{2j+1}^z )   \ , \label{kinetic}
\end{equation}
and

\begin{equation}
V=\sum_j {\bf S}_{2j-1}\cdot {\bf S}_{2j} \ . \label{perturb}
\end{equation}
We shall perform series expansions with respect to $\beta$.
The unperturbed states are

\begin{equation}
\vert \ \{ \alpha_j \} \ \rangle =
\bigotimes_j \vert \alpha_j \rangle_{2j,\ 2j+1} , \label{stateu}
\end{equation}
where $\vert \alpha_j \rangle_{2j,2j+1}$ is one of four eigenstates
of the operator $S_{2j}^x S_{2j+1}^x + S_{2j}^y S_{2j+1}^y
+ \lambda S_{2j}^z S_{2j+1}^z$:
$\vert S \rangle_{2j,2j+1}
= \frac{1}{\sqrt{2}} \left( \vert \uparrow \rangle_{2j}
\vert \downarrow \rangle_{2j+1} \
- \ \vert \downarrow \rangle_{2j} \vert \uparrow \rangle_{2j+1}
\right) $,
$\vert T_{+1} \rangle_{2j,2j+1}
= \vert \uparrow \rangle_{2j} \vert \uparrow \rangle_{2j+1} $,
$\vert T_0 \rangle_{2j,2j+1}
= \frac{1}{\sqrt{2}} \left( \vert \uparrow \rangle_{2j}
\vert \downarrow \rangle_{2j+1} \
+ \ \vert \downarrow \rangle_{2j}
\vert \uparrow \rangle_{2j+1} \right)$
and $\vert T_{-1} \rangle_{2j,2j+1}
= \vert \downarrow \rangle_{2j} \vert \downarrow \rangle_{2j+1} $.
For $\lambda >-1$, the unperturbed ground state is

\begin{equation}
\vert G \rangle
= \bigotimes_j \vert S \rangle_{2j,\ 2j+1} \ . \label{stateg}
\end{equation}
The ground state energy is expanded in a power series as

\begin{equation}
E(\lambda, \beta)
= \sum_{n=0} E^{(n)} (\lambda) \beta^n \label{energyE}
\ .
\end{equation}
The ``specific heat'' is obtained by

\begin{equation}
C(\lambda,\beta) =
\frac{\partial^2}{\partial \beta^2} E(\lambda,\beta)  \label{sh}
\ .
\end{equation}
We use the disorder operators defined per site

\begin{equation}
{\cal D}^\alpha = \frac{1}{M} \sum_j D^\alpha (j)
\ ,
\end{equation}
where $\alpha=xy$ and $z$, and $M$ is the number of sites.
To calculate series for the ``magnetization'' of the operators,
we simply add a magnetic field

\begin{equation}
H=T+ \beta V+h{\cal D}^\alpha.
\end{equation}
The ground state energy in the presence of $h$ is calculated
in a power series of $\beta$ and $h$ as

\begin{equation}
E(\lambda; \beta, h)
= \sum_{n=0} \sum_{m=0} E^{(n,m)}(\lambda) \beta^n h^m
\ \ .
\end{equation}
The expectation value of the ``magnetization'' is given by

\begin{equation}
\langle {\cal D}^{\alpha} \rangle =
\frac{\partial E(\lambda; \beta, h)}{\partial h} |_{h=0}
\ ,
\label{mag}
\end{equation}
and the ``susceptibility'' is

\begin{equation}
\frac{\partial \langle \ {\cal D}^\alpha \rangle}{\partial h} =
\frac{\partial^2 E(\lambda; \beta, h)}{\partial h^2} |_{h=0} \ .
\end{equation}

We use the Pad\'e method
to estimate critical points $\beta_c$ and critical indices.
The definition of critical indices are as follows:

\begin{equation}
\frac{\partial^2}{\partial \beta^2} E(\lambda,\beta)
\sim
{\rm constant} \left| \frac{\beta_c - \beta}{\beta_c}
\right|^{-\alpha} ,
\ \
(\beta \to \beta_c),
\end{equation}

\begin{equation}
\frac{\partial E(\lambda; \beta, h)}{\partial h} \vert_{h=0}
\sim
{\rm constant} \ \left| \frac{\beta_c - \beta}{\beta_c}
\right|^{-\beta_j} ,
\ \
(\beta \to \beta_c),
\end{equation}

\begin{equation}
\frac{\partial^2 E(\lambda; \beta, h)}{\partial h^2} \vert_{h=0}
\sim
{\rm constant} \ \left| \frac{\beta_c - \beta}{\beta_c}
\right|^{-\gamma_j} ,
\ \
(\beta \to \beta_c),
\end{equation}
where $j$ is $xy$ and $z$.

\subsection{Result from Series Analysis}

The quantities calculated by series expansions were
$\left( 1 \right)$ the ``specific heat'' (13 terms),
$\left( 2 \right)$ the ``magnetization'' (15 terms),
$\left( 3 \right)$ the ``susceptibility'' (13 terms).
Critical points and critical indices are evaluated
by the Dlog Pad\'e method \cite{dlog}.
The estimates for the quantities are obtained by averaging
the three or four highest-order diagonal elements
and near-diagonal elements of the Pad\'e tables.
Error bars are then set to include these three or four values.

\subsubsection{region \ $\beta >0$}
\label{region1}

For the region $-1<\lambda <1$,
the Haldane phase is characterized
by $\langle {\cal D}^{\alpha} \rangle$
$\ne 0$ ($\alpha=xy$ and $z$).
The disordered phase (E) is characterized
by $\langle {\cal D}^{\alpha} \rangle$ $=0$ ($\alpha=xy$ and $z$).
Thus the phase boundary (Line 1) between these phases is determined
from series for ${\cal D}^{\alpha}$ ($\alpha=xy$ and $z$).
Critical points obtained from series
for $\langle {\cal D}^{xy} \rangle$ and $\langle {\cal D}^z \rangle$
show well convergence.
We estimate critical points
from series for $\langle {\cal D}^{xy} \rangle$ (Fig.\ \ref{phase}).
In Fig.\ \ref{xybetap} and \ref{zbetap},
critical indices $\beta_j$'s ($j$ = $xy$ and $z$) are shown.
Results for $\gamma_j$'s ($j$ = $xy$ and $z$) and $\alpha$ are shown
in Fig.\ \ref{xygamma}, \ \ref{zgamma}
and \ \ref{alpha} respectively.
In this region, critical indices $\beta_j$'s ($j$ = $xy$ and $z$)
vary continuously,
and seem to be divergent near $\lambda =-1$.
Critical index $\beta_z$ agrees well with the result by Hida
obtained numerically \cite{Hida4}.
Critical indices $\gamma_j$'s ($j$ = $xy$ and $z$) and $\alpha$
are also varying continuously.
Critical index $\gamma_{xy}$
also seems to be divergent near $\lambda=-1$.
However, we cannot determine critical indices accurately
close to $\lambda =-1$ due to poor convergence.
In the case of the Ashkin-Teller quantum chain
or the staggered $XXZ$ model,
there is the ``critical fan'' in a finite region
of the parameter space
$-1 <\lambda < -\frac{1}{\sqrt{2}}$ \cite{KdK}.
The ``critical fan'' is a region
where a line of continuously varying criticality ``fans out''
and becomes an area of critical behavior.
In that case, critical indices show divergence
near $\lambda =-\frac{1}{\sqrt{2}}$ on the self-dual line
of the Ashkin-Teller quantum chain.
In the present model,
there is no sign of a ``critical fan'' within this analysis,
since critical points show well convergence
and critical indices does not show divergence.
However, we cannot discuss the existence of the ``critical fan''
near $\lambda=-1$ for poor convergence of critical indices.

For $\beta=1,\ \lambda=1$, the model is solvable
by the Bethe ansatz \cite{Bethe}.
We can map it to the six-vertex model at this point \cite{Bax}.
The extended scaling relations predict critical indices
$\alpha$, $\beta_j$'s and $\gamma_j$'s ($j$ = $xy$ and $z$),
with the help of the standard scaling relations.
These indices have been obtained from mappings
between the six-vertex model
and the Gaussian model \cite{KB,Kad2}.
The equivalent mapping is that between the $XXZ$ model
and the Tomonaga-Luttinger model \cite{LP}.
The estimated values of the critical point
from the series analysis are,
$\beta_c = 1.003 \pm 0.002 $ (from series for the ``specific heat''),
$\beta_c = 1.00168 \pm 0.00004 $
(from series for $\langle {\cal D}^{xy} \rangle$),
$\beta_c = 0.99782 \pm 0.00009 $
(from series for $\langle {\cal D}^{xy} {\cal D}^{xy} \rangle$ ).
The critical indices $\alpha$, $\beta_j$'s and $\gamma_j$'s
($j$ = $xy$ and $z$) are

\begin{equation}
\alpha = 0.55 \pm 0.01 \ \ \ ( \frac{2}{3}
= 0.666 \cdot \cdot \cdot \ ) \ ,
\end{equation}
\begin{equation}
\beta_j = 0.08202 \pm 0.00008 \ \ \
( \frac{1}{12} = 0.0833 \cdot \cdot \cdot \ ) \ ,
\end{equation}
\begin{equation}
\gamma_j = 1.23 \pm 0.02 \ \ \ ( \frac{7}{6}
= 1.166 \cdot \cdot \cdot \ ) \ .
\end{equation}
The values in parentheses are the expected values from the mappings.

Next we consider the region $\lambda > 1$.
The Haldane phase is characterized by
$\langle {\cal D}^{\alpha} \rangle \ne 0$ ($\alpha = xy$ and $z$).
The $S=1/2$ N\'eel phase (F) is characterized
by $\langle {\cal D}^{xy} \rangle = 0$
and $\langle {\cal D}^{z} \rangle \ne 0$.
Thus the phase boundary (Line 2) between these phases
is determined from series for ${\cal D}^{xy}$.
In Phase E,
both of $\langle {\cal D}^{xy} \rangle$
and $\langle {\cal D}^z \rangle$ vanish.
Therefore, the phase boundary (Line 3)
is determined from series for ${\cal D}^{z}$.
These results are shown in Fig.\ \ref{phase}.
Line 2 and Line 3 belong to the Ising model universality class
(see Sec.\ \ref{sec:phasedia}).
Thus critical indices $\alpha$, $\beta_{xy}$
and $\gamma_{xy}$ must be $0$, $\frac{1}{8}$
and $\frac{7}{4}$, respectively
on Line 2.
Critical indices $\alpha$, $\beta_{z}$
and $\gamma_{z}$ must be $0$, $\frac{1}{8}$
and $\frac{7}{4}$, respectively
on Line 3.
The series analysis shows
that Line 2 approaches the line $\beta_c =2$,
which is consistent with the result in Sec.\ \ref{sec:phasedia}.
For $\lambda >1$,
Critical index $\beta_{xy}$ agrees well with that of the Ising model
$\frac{1}{8}$.
Critical index $\gamma_{xy}$ also agrees well
with that of the Ising model
$\frac{7}{4}$.
Near $\lambda =1$ they do not agree with the Ising values.
We regard this as a numerical effect.
Critical index $\alpha$ does not agree well with the Ising value
$\alpha=0$ as $\lambda$ becomes large.
In the series analysis for $\langle {\cal D}^z \rangle$,
there is a pole near Line 2.
We choose the second pole to determine
the critical points and critical indices for Line 3
as shown in Fig.\ \ref{phase}.
The operator ${\cal D}^z$ has nonzero expectation value
in Phase A and in Phase F.
We regard the first pole as unphysical consequences,
since  series for ${\cal D}^z$ has no singularity on Line 2.
The series analysis shows
that Line 3 approaches the line $\beta=\frac{1}{2}\lambda$.
This is consistent with the result in Sec.\ \ref{sec:phasedia}.
However, their convergence is poor.
Critical indices $\alpha$, $\beta_z$ and $\gamma_z$
are not included here
for poor convergence.
Critical indices $\beta_z$ estimated from series analysis
are $\beta_z =0.03 \sim 0.1$ for $\lambda = 2 \sim 5$.
They do not agree well with the Ising values.
Critical index $\gamma_z$ does not converge.
We think that these are due to the first pole near Line 2
and large value of $\beta_c$.

\subsubsection{region \ $\beta <0$}

Let us consider the region $\lambda >0$.
The $S=1$ N\'eel phase (B) is characterized
by $\langle {\cal D}^{xy} \rangle =0$
and $\langle {\cal D}^{z} \rangle \ne 0$.
Thus the phase boundary (Line 4)
between the Haldane and the $S=1$ N\'eel phases
is determined from series for ${\cal D}^{xy}$.
The best estimates are obtained from series
for $\langle {\cal D}^{xy} \rangle$
as shown in Fig.\ \ref{phase}.
Line 4 belongs to the Ising model universality class
(see Sec.\ \ref{sec:phasedia}).
Therefore, critical indices are expected to take the Ising values
throughout Line 4.
In the limit $\beta \to -\infty$, Line 4 is expected to approach
the N\'eel-Haldane transition point $\lambda_1$
of the $S=1$ $XXZ$ model.
The critical points and critical indices obtained
from series for $\langle {\cal D}^{xy} \rangle$
are well convergent for $\lambda \stackrel{>}{\sim} 2.5$,
and stable for $2 \stackrel{<}{\sim} \lambda \stackrel{<}{\sim} 2.5$.
However, they are unstable for $\lambda \stackrel{<}{\sim} 2$.
No pole was found in every highest-order element of Pad\'e tables
for $0<\lambda \stackrel{<}{\sim} 1.2$
except for them with small residue ($< 10^{-5}$).
Note that if we investigate the behavior of Pad\'e tables carefully,
they show a highly unstableness near $\lambda \sim 1.8$.
Thus, for the region
$1.2 \stackrel{<}{\sim} \lambda \stackrel{<}{\sim} 1.8$,
we regard the critical points
as poor convergence of the Pad\'e approximation.
For large $\lambda$,
the critical line approaches the line $\beta_c =-2$,
which is consistent with the result in Sec.\ \ref{sec:phasedia}.
Critical index $\beta_{xy}$ is well convergent to
that of the Ising model $\frac{1}{8}$ (see Fig.\ \ref{xybetan}).
The result for $\gamma_{xy}$ is shown in Fig.\ \ref{xygamman}.
It shows a good agreement with the Ising value $\frac{7}{4}$.
Convergence of critical indices become poor around $\lambda =2$.
This is due to large value of $\beta_c$.

Next we focus on the region $-1<\lambda<0$.
In the $XY$-like phase, both of $\langle {\cal D}^{xy} \rangle$
and $\langle {\cal D}^{z} \rangle$ vanish.
Thus the phase boundary (Line 5)
between the Haldane and the $XY$-like phases
is determined from series for ${\cal D}^{\alpha}$
($\alpha = xy$ and $z$).
We choose the operator ${\cal D}^z$ to determine Line 5,
since critical points obtained from it converges better.
The critical points and indices are stable near the point
$(\lambda ,\ \beta)=(-1,\ 0)$.
The convergence of them becomes poor when $\lambda$ approaches $0$,
and it is difficult to determine the critical points
and critical indices.
Continuously varying critical exponents are observed
for $\beta_{xy}$ and $\beta_{z}$ along this line, which are shown
in Fig.\ \ref{xybetan} and Fig.\ \ref{zbetan} respectively.
Critical indices $\alpha$ and $\gamma_j$'s ($j$ = $xy$ and $z$)
are not included here for poor convergence.
These values estimated from series analysis are
$\alpha =2.2 \pm 1.1$, $\gamma_{xy} =3.4 \pm 2.6$
and $\gamma_z =3.1 \pm 2.6$ at $\lambda =-0.8$.
Line 5 is expected to be in the universality class
of the Kosterlitz-Thouless transition.
However, we cannot evaluate critical index $\eta$
for poor convergence of critical indices $\gamma_j$'s
($j$ = $xy$ and $z$).

\section{Summary and discussion}

We studied the ground state properties of the $S=1/2$ quantum spin
chain with bond alternation.
This model is equivalent to the $S=1$ antiferromagnetic $XXZ$ model
in the strong ferromagnetic coupling limit.
We focused on the disorder operators.
These operators are also obtained from the disorder operators
of the Ashkin-Teller quantum chain
by using the nonlocal unitary transformation.
The ``magnetization'' of these nonlocal operators measure
similar quantity
that is observed by the string order parameters
which is extended by Hida for the $S=1/2$ quantum spin chains.

The ground state phase diagram is obtained by series expansions.
Critical points are consistent with the phase diagram
which conjectured by Kohmoto and Tasaki \cite{KoT}
and determined numerically by Hida \cite{Hi3}.
In the narrow region $\lambda > -1$, $\beta \sim 0$,
it can be proved rigorously
that the ground state in the infinite volume limit is unique,
all the truncated correlation functions decay exponentially,
and there is a finite excitation gap \cite{KeT,KoT}.
The present study strongly suggest that this region
covers the wide area
surrounding by Lines 1, 2, 4 and 5 in Fig.\ \ref{phase}.
The results of critical indices show
that phase transition between the Haldane phase and the N\'eel phase
(Line 4 in Fig.\ \ref{phase}) belongs
to the Ising model universality class.
The Ising-like critical properties are originated
from the partial breakdown of the $Z_2 \times Z_2$ symmetry
in this case.
This supports that the hidden $Z_2 \times Z_2$ symmetry breaking
is to be a criterion to distinguish the Haldane gap system.
Although the nonlocal unitary transformation discussed here
is not exactly the same that used by Kohmoto and Tasaki,
the disorder operators have same critical properties with theirs.
This result also supports the proposal by Hida that the Haldane phase
which is characterized by the string order parameter
belongs to the same phase as that of the decoupled $S=1/2$ chain.

\acknowledgements
We wish to thank Kazuo Hida for useful discussions
and for sending his results prior to publication.

\figure{
Phase diagram of the alternating quantum spin chain
with the Hamiltonian (\ref{s2}).
Estimates of critical points by a series analysis
are shown with error bars.
For those without an error bar, the error is smaller
than the size of the plotted point.
\label{phase}
}

\figure{
Critical index $\beta_{xy}$
as a function of $\lambda$ on Line $1$ and $2$.
\label{xybetap}
}

\figure{
Critical index $\beta_z$
as a function of $\lambda$ on Line $1$.
\label{zbetap}
}

\figure{
Critical index $\gamma_{xy}$
as a function of $\lambda$ on Line $1$ and $2$.
\label{xygamma}
}

\figure{
Critical index $\gamma_z$
as a function of $\lambda$ on Line $1$.
\label{zgamma}
}

\figure{
Critical index $\alpha$
as a function of $\lambda$ on Line $1$ and $2$.
\label{alpha}
}

\figure{
Critical index $\beta_{xy}$
as a function of $\lambda $ on Line $4$ and $5$.
\label{xybetan}
}

\figure{
Critical index $\gamma_{xy}$
as a function of $\lambda $ on Line $4$.
\label{xygamman}
}

\figure{
Critical index $\beta_z$
as a function of $\lambda $ on Line $5$.
\label{zbetan}
}


\begin{references}
\bibitem[*]{address}Staying at M.I.T. until Sep. 30 1993.
\bibitem{Bethe}H. A. Bethe, Z. Phys. {\bf 71}, 205 (1931).
\bibitem{Hald}
F. D. M. Haldane, Phys. Lett. {\bf 93A}, 464 (1983);
Phys. Rev. Lett. {\bf 50}, 1153 (1983).
\bibitem{gapless}
P. W. Anderson, Phys. Rev. {\bf 86}, 694 (1952);
R. Kubo, Phys. Rev. {\bf 87}, 568 (1952);
J. des Cloizeaux and J. J. Pearson,
Phys. Rev. {\bf 128}, 2131 (1962).
\bibitem{exp}
W. J. L. Buyers, R. M. Morra, R. L. Armstrong, M. J. Hogan,
P. Gerlack and K. Hirakawa, Phys. Rev. Lett. {\bf 56}, 371 (1986);
M. Steiner, K. Kakurai, J. K. Kjems, D. Petitgrand
and R. Pynn, J. Appl. Phys. {\bf 61}, 3953 (1987);
J. P. Renard, M. Verdaguer, L. P. Regnault, W. A. C. Erkelens,
J. Rossat-Mignod, J. Ribas, W. G. Stirling and C. Vettier,
J. Appl. Phys. {\bf 63}, 3583 (1988);
K. Katsumata, H. Hori, T. Takeuchi, M. Date and A. Yamagishi
and J. P. Renard, Phys. Rev. Lett. {\bf 63}, 86 (1989);
Y. Ajiro, T. Goto, H. Kikuchi, T. Sakakibara and T. Inami,
Phys. Rev. Lett. {\bf 63}, 1424 (1989);
M. Hagiwara, K. Katsumata, I. Affleck, B. I. Halperin
and J. P. Renard, Phys. Rev. Lett. {\bf 65}, 3181 (1990).
\bibitem{num0}
R. Botet, R. Jullien and M. Kolb, Phys. Rev. {\bf B28}, 3914 (1983);
H. J. Schulz and T. Ziman, Phys. Rev. {\bf B33}, 6545 (1986).
\bibitem{num}
M. P. Nightingale and H. W. Bl$\ddot{o}$te,
Phys. Rev. {\bf B33}, 659 (1986);
M. Takahashi, Phys. Rev. Lett. {\bf 62}, 2313 (1989);
T. Sakai and M. Takahashi, Phys. Rev. {\bf B42}, 1090 (1990).
\bibitem{boundaryNH}K. Nomura, Phys. Rev. {\bf B40}, 9142 (1989).
\bibitem{num1}
K. Kubo and S. Takada, J. Phys. Soc. Jpn, {\bf 55}, 438 (1986);
F. C. Alcaraz and Y. Hatsugai, Phys. Rev. {\bf B46}, 13914 (1992).
\bibitem{theo}I. Affleck, T. Kennedy, E. H. Lieb and H. Tasaki,
Phys. Rev. Lett. {\bf 59}, 799 (1987);
Commun. Math. Phys. {\bf 115}, 477 (1988).
\bibitem{defect}
G. G\'omez-Santos, Phys. Rev. Lett. {\bf 63}, 790 (1989);
H. -J. Mikeska, Europhys. Lett. {\bf 19}, 39 (1992);
H. K$\ddot{o}$hler and R. Schilling,
J. Phys. Cond. Matter, {\bf 4}, 7899 (1992).
\bibitem{DN1}
M. den Nijs and K. Rommelse, Phys. Rev. {\bf B40}, 4709 (1989).
\bibitem{KeT}
T. Kennedy and H. Tasaki, Phys. Rev. {\bf B45}, 304 (1992);
H. Tasaki, Phys. Rev. Lett. {\bf 66}, 798 (1991).
\bibitem{GA}
S. M. Girvin and D. P. Arovas, Phys. Scr. {\bf T27}, 156 (1989);
Y. Hatsugai and M. Kohmoto, Phys. Rev. {\bf B44}, 11789 (1991).
\bibitem{T}
H. Tasaki (private communication);
M. Oshikawa, J. Phys. Cond. Matter, {\bf 4}, 7469 (1992);
Y. Hatsugai, J. Phys. Soc. Jpn, {\bf 61}, 3856 (1992).
\bibitem{alt}
T. Nakano and H. Fukuyama, J. Phys. Soc. Jpn. {\bf 49}, 1679 (1980);
J. Phys. Soc. Jpn. {\bf 50}, 2489 (1981);
K. Okamoto, H. Nishimori and Y. Taguchi,
J. Phys. Soc. Jpn. {\bf 55}, 1458 (1986);
S. Yoshida and K. Okamoto, J. Phys. Soc. Jpn. {\bf 58}, 4367 (1989);
and references there in.
\bibitem{Hi1}K. Hida, Phys. Rev. {\bf B45}, 2207 (1992).
\bibitem{Hi2}K. Hida, Phys. Rev. {\bf B46}, 8268 (1992).
\bibitem{Hi3}K. Hida, Comp. Phys. as a New Frontier
in Cond. Matt. Res. Proc. July 86 (1992);
(and private communication).
\bibitem{KoT}
M. Kohmoto and H. Tasaki, Phys. Rev. {\bf B46}, 3486 (1992)
\bibitem{s2nut}S. Takada and H. Watanabe,
J. Phys. Soc. Jpn. {\bf 61}, 39 (1992);
S. Takada, J. Phys. Soc. Jpn. {\bf 61}, 428 (1992).
\bibitem{KdK}M. Kohmoto, M. den Nijs and L. P. Kadanoff,
Phys. Rev. {\bf B24}, 5229 (1981).
\bibitem{KaK}L. P. Kadanoff and M. Kohmoto,
Nucl. Phys. {\bf B190}[FS3], 671 (1981).
\bibitem{AT}J. Ashkin and E. Teller, Phys. Rev. {\bf 64}, 178 (1943).
\bibitem{Kog2}For a review,
see J. B. Kogut, Rev. Mod. Phys. {\bf 51}, 659 (1979).
\bibitem{Pade}For an example see D. S. Gaunt and A. J. Guttmann,
in {\it Phase Transitions and Critical Phenomena},
edited by C. Domb and M. S. Green (Academic, London, 1974), Vol. 3.
\bibitem{KW}
L. P. Kadanoff and H. Ceva, Phys. Rev. {\bf B3}, 3918 (1971).
\bibitem{KB}
L. P. Kadanoff and A. C. Brown, Ann. Phys. {\bf 121}, 318 (1979).
\bibitem{Kada}L. P. Kadanoff, Phys. Rev. {\bf B22}, 1405 (1980).
\bibitem{KT}J. M. Kosterlitz and D. J. Thouless,
J. Phys. C {\bf 6}, 1181 (1973);
J. M. Kosterlitz, J. Phys. C {\bf 7}, 1046 (1974).
\bibitem{KDP}
E. H. Lieb, Phys. Rev. Lett. {\bf 18}, 692, 1046 (1967);
Phys. Rev. Lett. {\bf 19}, 108 (1967);
For a review see, E. H. Lieb and F. Y. Wu,
in {\it Phase Transitions and Critical Phenomena},
edited by C. Domb and M. S. Green (Academic, London, 1972), Vol. 1.
\bibitem{KK}
L. P. Kadanoff and M. Kohmoto, J. Phys. A {\bf 14}, 1291 (1981).
\bibitem{dlog}G. A. Baker Jr., Phys. Rev. {\bf 124}, 768 (1961).
\bibitem{Hida4}K. Hida (unpublished and private communication).
\bibitem{Bax}
R. J. Baxter, Phys. Rev. Lett. {\bf 26}, 832 (1971);
M. Barbar and R. J. Baxter, J. Phys. C {\bf 6}, 2913 (1973);
L. P. Kadanoff, Phys. Rev. Lett. {\bf 39}, 903 (1977).
\bibitem{Kad2}
L. P. Kadanoff, Ann. Phys. (N.Y.) {\bf 120}, 39 (1979);
H. J. F. Knops, Ann. Phys. (N.Y.) {\bf 128}, 448 (1981).
\bibitem{LP}
A. Luther and I. Peschel, Phys. Rev. {\bf B12}, 3908 (1975);
M. den Nijs, Phys. Rev. {\bf B23}, 6111 (1981);
J. L. Black and V. J. Emery, Phys. Rev. {\bf B23}, 429 (1981).




\end{references}
\end{document}